\documentclass[11pt]{article}

\def\mytitle#1{\setcounter{equation}{0}
\setcounter{footnote}{0}
\begin{flushleft}\Large\textbf{#1}\end{flushleft}
\vspace{0.25cm}}
\def\myname#1{\leftline{{\large #1}}\vspace{-0.13cm}}
\def\myplace#1#2{\small\begin{flushleft}\textit{#1}\\
\texttt{#2}\end{flushleft}}
\newenvironment{contribution}{\normalsize\noindent}{}
\def\myclassification#1{\small\noindent
Pacs no :
       #1\vspace{0.5cm}}
\usepackage{graphicx}
\begin{document}

\mytitle{GRAVITATIONAL COLLAPSE IN VAIDYA SPACE-TIME FOR GALILEON
GRAVITY THEORY}

\vskip0.2cm \myname{Prabir Rudra\footnote{prudra.math@gmail.com}}
\myplace{Department of Mathematics, Indian Institute of
Engineering Science and Technology, Shibpur, Howrah-711 103, India.\\
Department of Mathematics, Pailan College of Management and
Technology, Bengal Pailan Park, Kolkata-700 104, India.}{}

\vskip0.2cm \myname{Ujjal
Debnath\footnote{ujjaldebnath@yahoo.com}} \myplace{Department of
Mathematics, Indian Institute of Engineering Science and
Technology, Shibpur, Howrah-711 103, India.}{} \vskip0.2cm

\begin{abstract}
The motive of this work is to study gravitational collapse in
Vaidya space-time embedded in Galileon gravity theory. Galileon
gravity is in fact an infrared modification of Einstein gravity,
which was proposed as a generalization of the 4D effective theory
in DGP brane model. Vaidya's metric is used all over to follow the
nature of future outgoing radial null geodesics. Detecting whether
the central singularity is naked or wrapped by an event horizon,
by the existence of future directed radial null geodesic emitted
in past from the singularity is the basic objective. To point out
the existence of positive trajectory tangent solution, both
particular parametric cases (through tabular forms) and wide range
contouring process have been applied. Precisely, the EoS in
perfect fluid satisfies a wide range of phenomena: from dust to
exotic fluid like dark energy. We have used the EoS parameter $k$
to determine the end collapse state in different cosmological era.
\end{abstract}

\myclassification{ 04.20.Dw, 98.65.D}

\section{INTRODUCTION}
Right from the discovery of Newton's theory of gravitation, one
question that has been haunting the human mind is `how gravity
works at cosmological distances?' Although there was no definite
answer to this question, yet there was a feeling that the story at
large distances is quite different to Newton's gravitation. With
the discovery of the cosmic acceleration in the late twentieth
century, this feeling started transforming into belief. As a
possible explanation for the cosmic acceleration, modifications to
gravity at cosmological distances have recently received much
attention. Numerous modified gravity theories have been proposed
during the last decade, some of which worth mentioning are, loop
Quantum gravity (LQG) \cite{Ashtekar1, Rovelli1}, Gauss-Bonnet
gravity \cite{Carroll1, Nojiri1}, scalar-tensor theories
\cite{Hirano1, Hirano2, Hartnett1}, $f(R)$ gravity \cite{Felice1},
DGP braneworld model \cite{Dvali1}, Galileon gravity
\cite{Nicolis1, Deffayet1, Deffayet2, Chow1, Silva1}, etc.

Modification to gravity can be done keeping in mind that the
deviations from General Relativity (GR) can only be allowed at
large distances. At close distances the theory should coincide
with Newton's theory of gravitation thus ensuring the consistency
of the model with the solar system experiments. We know that the
DGP brane model admits a self-accelerating solution (cosmic
acceleration in the absence of any matter in the universe)
\cite{Deffayet3}. But unfortunately this solution is plagued with
ghost instabilities, which renders the model rather invalid
\cite{Koyama1}. There are many other modified gravity models
suffering from this instability problem \cite{Koyama2}.

Of late an infrared modification of classical gravitation was
proposed, as a generalization of the 4D effective theory in the
DGP model \cite{Nicolis1}. The theory considers a self-interaction
term of the form $\left(\nabla \phi\right)^{2}~^{\fbox{}}~\phi$ in
order to recover GR in high density regions. The most striking
feature of the theory is that it is invariant under the Galileon
shift symmetry, $\delta_{\mu}\phi\rightarrow
\delta_{\mu}\phi+c_{\mu}$, in the Minkowski background. Due to
this invariance the equation of motion remains a second order
differential equation, preventing the introduction of extra
degrees of freedom, which are usually associated with
instabilities.

The study of gravitational collapse was started by Oppenheimer and
Snyder in 1939 \cite{Openheimer}. In classical GR, the
gravitational collapse is a problem of great curiosity as we can
get at least two types of singularities from it. One, covered by
an event horizon, is coined as a black hole (BH) whereas the
singularity alone is popular as Naked Singularity (NS). Now, one
would always like to test the validity of cosmic censorship
hypothesis (CCH) laid down by Penrose, \cite{Penrose} which stated
that the end result of a collapse is bound to be a singularity
covered by an event horizon, i.e., BH. In last few decades there
have been extensive research \cite{Eardly1, Chris1, Newman1,
Dwivedi1, Waugh1, Ori1} where the possibility of formation of NS
has been investigated. To determine the end state of collapse (NS
or BH), the Vaidya solution \cite{Vaidya1} is utilized on many
occasions. Harko et al \cite{Harko1} have also
studied the gravitational collapse in Vaidya space-time.\\

Gravitational collapse has been extensively studied in modified
gravity theory i.e., in Gauss-Bonnet, $f(R)$ gravity, Lovelock
theory etc \cite{Maeda1,Jhingan1,Ghosh1,Ghosh2,Joshi_rev2,
Singh_rev, Joshi_rev, Joshi1, Lake1, Szekeres1, Joshi2, Patil1,
Patil2, Debnath1, Debnath2, Banerjee1, Rudra1, Debnath3, Rudra}.
Scheel \cite{Scheel1} demonstrated that Openheimer-Snyder collapse
in Brans-Dicke theory results in BHs rather than NSs with the
positive values of brans-dicke parameter $\omega$, they have
speculated that the apparent horizon of a BH can pass outside the
event horizon causing the decrease in the surface area over time.
The values of $\omega$ forces the BH, to radiate its scalar mass
to infinity soon after the initial collapse. In \cite{Dong1} we
get another relevant and interesting work regarding the
gravitational collapse in the background of Brans-Dicke theory of
gravity discussing the effects of different values of $\omega$.
Rudra et al in \cite{Rudra2} has shown that lower the value of
$\omega$, greater is the chance of getting an NS. \\

In 1996, Husain \cite{{Husain1}} has studied the non-static
spherically symmetric solutions of Einstein equations, for a null
fluid source, where the density $\rho$ and pressure $p$ of the
fluid is related by barotropic equation of state $p=k\rho$. All of
the new solutions supported the cosmic censorship conjecture.
Later the Vaidya solution was generalized by Wang et al
\cite{Wang1}. Moreover the Husain solution has been extensively
used to study the formation of a black hole with short hair
\cite{Brown1}. The most recent development in Husain solution was
witnessed when the gravitational collapse of the Husain solution
in four and five dimensional space-times was studied by Patil et
al \cite{Patil1}.

Keeping all the previous works of gravitational collapse in GR/
different gravity theories in mind we feel it will be of a great
interest if we investigate the existence of radial null geodesic
from the collapsing body in the back ground of Galileon gravity
theory. The scalar factor present in the theory may help in
collapse to form a NS more prominently than the GR does. As in
\cite{Rudra2} we will check that whether the nature of the end
state of collapse can be regulated by any parameter of Galileon
gravity. In this concern we must recall the fact that in
\cite{Rudra1} while working with Lovelock gravity, we saw that
greater the deviation from Einstein gravity greater was the
tendency to have the NS. So in this paper, we are mainly studying
the nature of singularities (BH or NS) formed by the gravitational
collapse in Galileon gravity and the conditions governing these.
In section (\ref{calculation}), we present the brief overview of
generalized Vaidya solution in Galileon gravity. We will first
construct the Einstein field equations in Galileon theory for the
Vaidya metric and then with a proper choice of the structural
dependence of the potential term upon the scalar field we will
determine the $m(t, r)$, the mass term. In the next two sections
we investigate the behaviour/existence of the outgoing radial null
geodesic from the singularity taking the Vaidya metric with the
mass term $m(t, r)$ derived in the last section. Finally, the
paper ends with some concluding remarks in section (\ref{Discussion}).\\

\section{Field Equations and the Solutions in Vaidya Space-Time in the
background of Galileon Gravity}\label{calculation}

The Galileon theory is described by the action \cite{Nicolis1,
Deffayet1, Deffayet2, Chow1, Silva1}:
\begin{equation}\label{Lag}
S=\int d^{4} x \sqrt{-g}\left[\phi R- \frac{\omega}{\phi}
\left(\nabla \phi\right)^{2}+ f(\phi)^{\fbox{}}~\phi \left(\nabla
\phi\right)^{2}+{\cal L}_{m}\right]
\end{equation}
where where $\phi$ is the Galileon field and the coupling
$f(\phi)$ has dimension of length, $\left(\nabla
\phi\right)^{2}=g^{\mu\nu}\nabla_{\mu}\phi \nabla_{\nu}\phi$,~~~~
$^{\fbox{}}~\phi=g^{\mu\nu}\nabla_{\mu}\nabla_{\nu}\phi$~~ and~
${\cal L}_{m}$~~ is the matter Lagrangian. Here we consider the
metric in spherically symmetric space-time in the form
\cite{Vaidya1}
\begin{equation}\label{collapse2.1}
ds^{2}=-\left(1-\frac{m(t,r)}{r}\right)dt^{2}+2dtdr+r^{2}d\Omega_{2}^{2}
\end{equation}
where $r$ is the radial co-ordinate and $t$ is the null
co-ordinate, $m(t,~r)$ gives the gravitational mass inside the
sphere of radius $r$ and $d\Omega_{2}^{2}$ is the line element on
a unit 2-sphere.

Variation with respect to the metric gives the Einstein's
equations,

$$G_{\mu
\nu}=\frac{T_{\mu\nu}}{2\phi}+\frac{1}{\phi}\left(\nabla_{\mu}\nabla_{\nu}\phi-g_{\mu\nu}~^{\fbox{}}~\phi\right)
+\frac{\omega}{\phi^{2}}\left[\nabla_{\mu}\phi\nabla_{\nu}\phi-\frac{1}{2}g_{\mu\nu}\left(\nabla\phi\right)^{2}\right]$$
\begin{equation}
-\frac{1}{\phi}\left\{\frac{1}{2}g_{\mu\nu}\nabla_{\lambda}[f(\phi)\left(\nabla
\phi\right)^{2}]\nabla^{\lambda}\phi-\nabla_{\mu}[f(\phi)\left(\nabla
\phi\right)^{2}]\nabla_{\nu}\phi+f(\phi)\nabla_{\mu}\phi\nabla_{\nu}\phi~^{\fbox{}}~\phi\right\}
\end{equation}

Now we consider two types of fluids like Vaidya null radiation and
a perfect fluid having the form of the energy-momentum tensor
\cite{Rudra1}
\begin{equation}\label{collapse2.3}
T_{\mu\nu}=T_{\mu\nu}^{(n)}+T_{\mu\nu}^{(m)}
\end{equation}
with
\begin{equation}\label{collapse2.4}
T_{\mu\nu}^{(n)}=\sigma l_{\mu}l_{\nu}
\end{equation}
and
\begin{equation}\label{collapse2.5}
T_{\mu\nu}^{(m)}=(\rho+p)(l_{\mu}\eta_{\nu}+l_{\nu}\eta_{\mu})+pg_{\mu\nu}
\end{equation}
where, $\rho$ and $p$ are the energy density and pressure for the
perfect fluid and $\sigma$ is the energy density corresponding to
Vaidya null radiation. The two eigen vectors of energy-momentum
tensor namely $l_{\mu}$ and $\eta_{\mu}$ are linearly independent
future pointing null vectors having components
\begin{equation}\label{collapse2.6}
l_{\mu}=(1,0,0,0)~~~~ and~~~~
\eta_{\mu}=\left(\frac{1}{2}\left(1-\frac{m}{r}\right),-1,0,0
\right)
\end{equation}
and they satisfy the relations
\begin{equation}\label{collapse2.7}
l_{\lambda}l^{\lambda}=\eta_{\lambda}\eta^{\lambda}=0,~
l_{\lambda}\eta^{\lambda}=-1
\end{equation}

The Einstein field equations ($G_{\mu\nu}=T_{\mu\nu}$) for the
metric (\ref{collapse2.1}) and the wave equation for the Galileon
field $\phi$ are the following:\\
\vspace{3mm}

~~~~~~~~~~~~~~~~~~~~~~~~~~~~\textbf{From $G_{00}=T_{00}$ we get,}

$$\frac{\left(r-m\right)m'+r\dot{m}}{r^{3}}=\frac{\sigma+\rho\left(1-\frac{m}{r}\right)}
{2\phi}+\frac{1}{\phi}\left[\ddot{\phi}-\left(\frac{m}{2r^{2}}-\frac{m'}{2r}\right)\dot{\phi}
-\left(\frac{m}{2r^{2}}-\frac{m^{2}}{2r^{3}}-\frac{m'}{2r}+\frac{mm'}{2r^{2}}
+\frac{\dot{m}}{2r}\right)\phi'\right.$$

$$\left.+\left(1-\frac{m}{r}\right)\left\{2\dot{\phi}'-\phi'\left(\frac{m'}{r}
-\frac{3m}{r^{2}}+\frac{2}{r}\right)+\left(1-\frac{m}{r}\right)\phi''\right\}\right]
+\frac{\omega}{\phi^{2}}\left[\dot{\phi}^{2}+\frac{1}{2}\left(1-\frac{m}{r}\right)
\phi'\left(2\dot{\phi}+\left(1-\frac{m}{r}\right)\phi'\right)\right]$$
\begin{equation}\label{1}
+\frac{1}{\phi}\left[\frac{1}{2}\left(1-\frac{m}{r}\right)\left\{\phi'\nabla_{0}U
+\left(\dot{\phi}+\left(1-\frac{m}{r}\right)\phi'\right)\nabla_{1}U\right\}
-\dot{\phi}\nabla_{0}U+f(\phi)\dot{\phi}^{2}\left\{2\dot{\phi}'-\phi'\left(\frac{m'}{r}
-\frac{3m}{r^{2}}+\frac{2}{r}\right)\left(1-\frac{m}{r}\right)\phi''\right\}\right]~,
\end{equation}
\vspace{3mm}

~~~~~~~~~~~~~~~~~~~~~~~~~~~~~~~\textbf{$G_{11}=T_{11}$ gives,}

$$\frac{\phi''}{\phi}+\frac{\omega\phi'^{2}}{\phi^{2}}-\frac{1}{\phi}\left[-f(\phi)
\left\{2\phi'^{2}\dot{\phi}'+2\dot{\phi}\phi'\phi''+2\left(1-\frac{m}{r}\right)
\phi'^{2}\phi''+\frac{m}{r^{2}}\phi'^{3}\right\}-f'(\phi)\left\{2\phi'^{3}\dot{\phi}
+\left(1-\frac{m}{r}\right)\phi'^{4}\right\}\right.$$
\begin{equation}\label{2}
\left.+f(\phi)\phi'^{2}\left\{2\dot{\phi}'-\phi'\left(\frac{m'}{r}-\frac{3m}{r^{2}}
+\frac{2}{r}\right)+\left(1-\frac{m}{r}\right)\phi''\right\}\right]=0,
\end{equation}
\vspace{3mm}

~~~~~~~~~~~~~~~~~~~~~~~~~~~~~~~~~\textbf{$G_{01}=T_{01}$ gives,}

$$\frac{m'}{r^{2}}=\frac{\rho}{2\phi}+\frac{1}{\phi}\left[\dot{\phi}'+\phi'\left(\frac{m'}{2r}
-\frac{m}{2r^{2}}\right)-\phi'\left(\frac{m'}{r}-\frac{3m}{r^{2}}+\frac{2}{r}\right)
+\phi''\left(1-\frac{m}{r}\right)\right]+\frac{\omega}{2\phi^{2}}\left(1-\frac{m}{r}\right)\phi'^{2}$$

\begin{equation}\label{3}
+\frac{1}{\phi}\left[\frac{1}{2}\nabla_{0}U\left(\dot{\phi}-2\phi'\right)+\frac{1}{2}\phi'\nabla_{1}U
+f(\phi)\dot{\phi}\phi'\left\{2\dot{\phi}'-\phi'\left(\frac{m'}{r}-\frac{3m}{r^{2}}+\frac{2}{r}\right)
+\left(1-\frac{m}{r}\right)\phi''\right\}\right]~,
\end{equation}

~~~~~~~~~~~~~~~~~~~~~~~~~~~~~~~~~~~\textbf{$G_{22}=T_{22}$
gives,}

$$\frac{1}{2}rm''=\frac{\omega}{\phi^{2}}\left[\frac{r^{2}}{2}\phi'\left\{2\dot{\phi}
+\left(1-\frac{m}{r}\right)\phi'\right\}\right]-\frac{1}{\phi}\left[r^{2}\left\{\phi'\left(\frac{m'}{r}
-\frac{3m}{r^{2}}+\frac{2}{r}\right)-\left(1-\frac{m}{r}\right)\phi''-2\dot{\phi}'\right\}\right]$$
\begin{equation}\label{4}
+\frac{r^{2}}{2\phi}\left(\nabla_{0}U\dot{\phi}+\nabla_{1}U\phi'\right)-\frac{p
r^{2}}{2\phi}
\end{equation}
 and
\vspace{3mm}

~~~~~~~~~~~~~~~~~~~~~~~~~~~~~~~~~~~~~~\textbf{$G_{33}=T_{33}$
gives,}

$$\frac{pr^{2}}{2\phi}+\frac{1}{\phi}\left[r\dot{\phi}-\left(m-r\right)\phi'-r^{2}\left\{2\dot{\phi}'
-\phi'\left(\frac{m'}{r}-\frac{3m}{r^{2}}+\frac{2}{r}\right)+\left(1-\frac{m}{r}\right)\phi''\right\}\right]
-\frac{\omega}{\phi^{2}}\left[\frac{\phi'}{2}r^{2}\left(2\dot{\phi}+\phi'\left(1-\frac{m}{r}\right)\right)\right]$$
\begin{equation}\label{5}
-\frac{1}{2\phi}r^{2}\left(\dot{\phi}\nabla_{0}U+\phi'\nabla_{1}U\right)+\frac{1}{2}rm''=0
\end{equation}
where an over-dot and dash stand for differentiation with respect
to $t$ and $r$ respectively. Here $U=f(\phi)\left(\nabla
\phi\right)^{2}$ and correspondingly the expressions for
$\nabla_{0}U$ and $\nabla_{1}U$ are given as below:
\begin{equation}
\nabla_{0}U=f(\phi)\left[2\phi'\ddot{\phi}+2\dot{\phi}\dot{\phi}'
+\left(1-\frac{m}{r}\right)2\phi'\dot{\phi}'-\phi'^{2}\frac{\dot{m}}{r}\right]
+f'(\phi)\left[2\phi'\dot{\phi}^{2}+\left(1-\frac{m}{r}\right)\phi'^{2}\dot{\phi}\right],
\end{equation}

\begin{equation}
\nabla_{1}U=f(\phi)\left[2\phi'\dot{\phi}'+2\dot{\phi}\phi''+2\left(1-\frac{m}{r}\right)\phi'\phi''+\frac{m}{r^{2}}\phi'^{2}\right]+f'(\phi)\left[2\phi'^{2}\dot{\phi}+\left(1-\frac{m}{r}\right)\phi'^{3}\right]
\end{equation}

Due to the complicated nature of the field equations we cannot
solve them directly and get the expression for $\phi$. Therefore
we assume
\begin{equation}\label{7}
\phi(r,t)=P(r)Q(t)
\end{equation}
where $P(r)$ is an arbitrary function of $r$ and $Q(t)$ is an
arbitrary function of $t$. Since $f(\phi)$ is an arbitrary
function of $\phi$, \textbf{so in order to facilitate
calculations, we choose,}
\begin{equation}
f(\phi)=f_{0}\phi^{-2}
\end{equation}
where, $f_{0}$ is a constant. We assume the matter fluid obeys the
barotropic equation of state
\begin{equation}\label{8}
p=k\rho,~~~(k,~a~constant)
\end{equation}

Using  equations (11), (12), (14), (15), (16) and (17) we have the
solution for $Q(t)$ as,
\begin{equation}
Q(t)=\alpha_{1}e^{-\lambda t}
\end{equation}
 where $\alpha_{1}$ and $\lambda$ are arbitrary constants. Since
 no solution for $P(r)$ could be obtained due to the highly
 complicated nature of the field equations, we consider,
\begin{equation}
P(r)=\alpha r^{n}
\end{equation}
where $\alpha$ and $n$ are arbitrary constants. Now using the
above values of $P$ and $Q$ in the field equations we finally
arrive at a differential equation in $m(t,r)$ as,

\begin{equation}
r^{2}m''+\left[k+n\left(2+k\right)\right]rm'+\left[n\left\{2\left(k+1\right)\left(n-1\right)
-\left(5k+6\right)\right\}\right]m+2n\left[\left(3-n\right)\left(k+1\right)r+\left(\omega+k+2\right)\lambda
r^{2}\right]=0
\end{equation}

Solving the above differential equation we obtain the explicit
solution for $m$ as,
\begin{equation}\label{10}
m(t,r)=f_{1}(t)r^{\omega_{1}}+f_{2}(t)r^{\omega_{2}}+\frac{2n\left(3-n\right)\left(k+1\right)}{\left(1-\omega_{1}\right)\left(1-\omega_{2}\right)}r+\frac{2n\lambda\left(\omega+k+2\right)}{\left(2-\omega_{1}\right)\left(2-\omega_{2}\right)}r^{2}
\end{equation}
where
\begin{equation}\label{11}
\omega_{1},
\omega_{2}=\frac{\left[1-k-n\left(2+k\right)\right]\pm\sqrt{\left\{k+n\left(2+k\right)-1\right\}^{2}-4n\left\{2\left(k+1\right)\left(n-1\right)-\left(5k+6\right)\right\}}}{2}
\end{equation}
Here $f_{1}(t)$ and $f_{2}(t)$ are arbitrary functions of $t$.

Therefore the metric (\ref{collapse2.1}) can be written as

\begin{equation}\label{12}
ds^{2}=\left[-1+f_{1}(t)r^{\omega_{1}-1}+f_{2}(t)r^{\omega_{2}-1}+\frac{2n\left(3-n\right)\left(k+1\right)}{\left(1-\omega_{1}\right)\left(1-\omega_{2}\right)}+\frac{2n\lambda\left(\omega+k+2\right)}{\left(2-\omega_{1}\right)\left(2-\omega_{2}\right)}r\right]dt^{2}+2dtdr+r^{2}d\Omega_{2}^{2}
\end{equation}

which is called the Generalized Vaidya metric in Galileon gravity.

\section{Collapse Study}

We shall discuss the existence of NS in generalized Vaidya
space-time by studying radial null geodesics. In fact, we shall
examine whether it is possible to have outgoing radial null
geodesics which were terminated in the past at the central
singularity $r=0$. The nature of the singularity (NS or BH) can be
characterized by the existence of radial null geodesics emerging
from the singularity. The singularity is at least locally naked if
there exist such geodesics and if no such geodesics exist it is a
BH.

Let $R(t, ~r )$ is the physical radius at time $t$ of the shell
labelled by $r$. At the starting epoch $t=0$ we have assumed the
scaling (freedom) $R(0,~r)=r$. Now if there are future directed
radial null geodesics coming out of the singularity, with a well
defined tangent at  the singularity $\frac{dR}{dr}$ must tend to a
finite limit in the limit of approach to the singularity in the
past along these trajectories.

The point $(t_0, ~r)=(0,0)$ occurs, where the singularity $R(t_0,
0)=0$ occurs corresponds to the physical situation where matter
shells are crushed to zero radius. This kind of singularity
($r=0$) is known to be a central shell focusing singularity. The
singularity is a NS if there are future directed non-space like
curves in the space time with their past end points at the
singularity. Now if the outgoing radial null geodesics are to
terminate in the past at the central singularity at $r=0$ at
$t=t_0$ where $R(t_0, 0)=0$, then along these geodesics we should
have \cite{Singh_rev} $R\rightarrow 0$ as $r\rightarrow 0$.

The equation for outgoing radial null geodesics can be obtained
from equation (\ref{collapse2.1}) by putting $ds^{2}=0$ and
$d\Omega_{2}^{2}=0$ as
\begin{equation}\label{collapse2.24}
\frac{dt}{dr}=\frac{2}{\left(1-\frac{m(t,r)}{r}\right)}.
\end{equation}
Now we consider the time $t=0$, when the singularity forms at the
centre $r=0$. So we can recalled that $r=0,~t=0$ corresponds to a
central singularity. It can be seen easily that $r=0,~t=0$
corresponds to a singularity of the above differential equation.
Suppose $X=\frac{t}{r}$ then we shall study the limiting behavior
of the function $X$ as we approach the singularity at $r=0,~t=0$
along the radial null geodesic. If we denote the limiting value by
$X_{0}$ then
\begin{eqnarray}\label{collapse2.25}
\begin{array}{c}
X_{0}\\\\
{}
\end{array}
\begin{array}{c}
=lim~~ X \\
\begin{tiny}t\rightarrow 0\end{tiny}\\
\begin{tiny}r\rightarrow 0\end{tiny}
\end{array}
\begin{array}{c}
=lim~~ \frac{t}{r} \\
\begin{tiny}t\rightarrow 0\end{tiny}\\
\begin{tiny}r\rightarrow 0\end{tiny}
\end{array}
\begin{array}{c}
=lim~~ \frac{dt}{dr} \\
\begin{tiny}t\rightarrow 0\end{tiny}\\
\begin{tiny}r\rightarrow 0\end{tiny}
\end{array}
\begin{array}{c}
=lim~~ \frac{2}{\left(1-\frac{m(t,r)}{r}\right)} \\
\begin{tiny}t\rightarrow 0\end{tiny}~~~~~~~~~~~~\\
\begin{tiny}r\rightarrow 0\end{tiny}~~~~~~~~~~~~
 {}
\end{array}
\end{eqnarray}
Using equations (\ref{10}) and (\ref{collapse2.25}), we have
\begin{eqnarray}
\frac{2}{X_{0}}=
\begin{array}llim\\
\begin{tiny}t\rightarrow 0\end{tiny}\\
\begin{tiny}r\rightarrow 0\end{tiny}
\end{array}\left[1-f_{1}(t)r^{\omega_{1}-1}-f_{2}(t)r^{\omega_{2}-1}-
\frac{2n\left(3-n\right)\left(k+1\right)}{\left(1-\omega_{1}\right)
\left(1-\omega_{2}\right)}-\frac{2n\lambda\left(\omega+k+2\right)}
{\left(2-\omega_{1}\right)\left(2-\omega_{2}\right)}\frac{r}{t}\right]
\end{eqnarray}

Now choosing $f_{1}(t)=\delta t^{-(\omega_{1}-1)}$ ~and
~$f_{2}(t)=\epsilon t^{-(\omega_{2}-1)}$,~($\delta$ and $\epsilon$
are constants), we obtain the algebraic equation of $X_{0}$ as

\begin{equation}
\delta X_{0}^{2-\omega_{1}}+\epsilon
X_{0}^{2-\omega_{2}}-\left[1-\frac{2n\left(3-n\right)\left(k+1\right)}
{\left(1-\omega_{1}\right)\left(1-\omega_{2}\right)}\right]X_{0}+
2\left[1+\frac{n\lambda\left(\omega+k+2\right)}{\left(2-\omega_{1}\right)
\left(2-\omega_{2}\right)}\right]=0
\end{equation}
Outgoing radial null geodesic exists if the value of the tangent
near the singularity is positive i.e., $X_{0}>0$. Now if we get
only non-positive solution of the equation we can assure the
formation of a BH. Getting a positive root indicates a chance to
get a NS. Since the obtained equation is a highly complicated one,
it is extremely difficult to find out an analytic solution of
$X_{0}$ in terms of the variables involved. So our idea is to find
out different numerical solutions of $X_{0}$, by assigning
particular numerical values to the associated constants.

The different solutions of $X_{0}$ for different sets of
parametric values of ($\omega, \lambda, \delta, \epsilon, n$) and
for different stages of EoS parameter $k$,
which are given here in a tabular form (Table 1a-e).\\

\begin{center}
Table1a
\begin{tabular}{|l|}
\hline\hline\\ ~~~~~~~~~~~~~~~~~~~~~~~~~~~~~~~~~~~~~~~~~~~~~~~~For $k=1$ (stiff perfect fluid)\\
\hline\hline
~~~~$\omega$~~~~~~~~~~~$\lambda$~~~~~~~~~~~$\delta$~~~~~~~~~~~$\epsilon$~~~~~~~~~~~$n$
~~~~~~~~~~~~~~~Positive roots ($X_{0}$)
\\ \hline
\\
~~~~3~~~~~~~~~~~1~~~~~~~~~~~1~~~~~~~~~~~1~~~~~~~~~~~~1~~~~~~~~~~~~~~~~~~~~~~~~$-$
\\
~~~~3~~~~~~~~~~~1~~~~~~~~~~~1~~~~~~~~~~~1~~~~~~~~~~~~2~~~~~~~~~~~~~~~~~~~~~~~~~$-$
\\
~~~~3~~~~~~~~~~~1~~~~~~~~~~~1~~~~~~~~~~~1~~~~~~~~~~~~4~~~~~~~~~~~~~~~~~~~~~~$0.751311$
\\

\\ \hline \\
~~~~2~~~~~~~~~~~3~~~~~~~~~~0.1~~~~~~~~~0.2~~~~~~~~~~~1~~~~~~~~~~~~~~~~~~~~~~~~~$-$
\\
~~~~2~~~~~~~~~~~3~~~~~~~~~~0.1~~~~~~~~~0.2~~~~~~~~~~~2~~~~~~~~~~~~~~~~~~~~~~$0.864566$
\\
~~~~2~~~~~~~~~~~3~~~~~~~~~~0.1~~~~~~~~~0.2~~~~~~~~~~~4~~~~~~~~~~~~~~~~~~~~~~$0.911302$
\\

\\ \hline \\
~~~~1~~~~~~~~~~0.1~~~~~~~~~~2~~~~~~~~~~~1~~~~~~~~~~~~1~~~~~~~~~~~~~~~~~~~~~~~~~$-$
\\
~~~~1~~~~~~~~~~0.1~~~~~~~~~~2~~~~~~~~~~~1~~~~~~~~~~~~2~~~~~~~~~~~~~~~~~~~~~~~~~$-$
\\
~~~~1~~~~~~~~~~0.1~~~~~~~~~~2~~~~~~~~~~~1~~~~~~~~~~~~4~~~~~~~~~~~~~~~~~~~~~~$0.690284$

\\ \hline \\
~~~-2~~~~~~~~~~0.1~~~~~~~~~0.1~~~~~~~~~0.1~~~~~~~~~~~1~~~~~~~~~~~~~~~~~~~~~~~~~$-$
\\
~~~~2~~~~~~~~~~0.1~~~~~~~~~0.1~~~~~~~~~0.1~~~~~~~~~~~2~~~~~~~~~~~~~~~~~~~~~~$0.91185$
\\
~~~~2~~~~~~~~~~0.1~~~~~~~~~0.1~~~~~~~~~0.1~~~~~~~~~~~4~~~~~~~~~~~~~~~~~~~~~~$0.937988$

\\ \hline \\
~~~-3~~~~~~~~~~0.1~~~~~~~~~0.1~~~~~~~~~~1~~~~~~~~~~~~1~~~~~~~~~~~~~~~~~~~~~~~~~$-$
\\
~~~~3~~~~~~~~~~0.1~~~~~~~~~0.1~~~~~~~~~~1~~~~~~~~~~~~2~~~~~~~~~~~~~~~~~~~~~~$0.761759$
\\
~~~~3~~~~~~~~~~0.1~~~~~~~~~0.1~~~~~~~~~~1~~~~~~~~~~~~4~~~~~~~~~~~~~~~~~~~~~~$0.846213$
\\ \hline \\
~~~-4~~~~~~~~~~0.1~~~~~~~~~0.1~~~~~~~~~~1~~~~~~~~~~~~1~~~~~~~~~~~~~~~~~~~~~~~~~$-$
\\
~~~-4~~~~~~~~~~0.1~~~~~~~~~0.1~~~~~~~~~~1~~~~~~~~~~~~2~~~~~~~~~~~~~~~~~~~~~~$0.761758$
\\
~~~-4~~~~~~~~~~0.1~~~~~~~~~0.1~~~~~~~~~~1~~~~~~~~~~~~4~~~~~~~~~~~~~~~~~~~~~~$0.846213$
\\
\hline\hline\\
\end{tabular}

\end{center}
\begin{center}
Table1b
\begin{tabular}{|l|}

\hline\hline\\ ~~~~~~~~~~~~~~~~~~~~~~~~~~~~~~~~~~~~~~~~~~~~~~~~For $k=1/3$ (radiation)\\
\hline\hline
~~~~$\omega$~~~~~~~~~~~$\lambda$~~~~~~~~~~~$\delta$~~~~~~~~~~~$\epsilon$~~~~~~~~~~~$n$
~~~~~~~~~~~~~~~Positive roots ($X_{0}$)
\\
\hline\\
~~~~2~~~~~~~~~0.1~~~~~~~~~~~0.1~~~~~~~~~1~~~~~~~~~~~~1~~~~~~~~~~~~~~~~~~~~~~~~~~~~$1.24846$
\\
~~~~2~~~~~~~~~0.1~~~~~~~~~~~0.1~~~~~~~~~1~~~~~~~~~~~~2~~~~~~~~~~~~~~~~~~~~~~~~~~~~$0.688704$
\\
~~~~2~~~~~~~~~0.1~~~~~~~~~~~0.1~~~~~~~~~1~~~~~~~~~~~~4~~~~~~~~~~~~~~~~~~~~~~~~~~~~$0.818295$
\\ \hline \\
~~~~1~~~~~~~~~0.1~~~~~~~~~~~~2~~~~~~~~~0.1~~~~~~~~~~~1~~~~~~~~~~~~~~~~~~~~~~~~~~~~$0.691721$
\\
~~~~1~~~~~~~~~0.1~~~~~~~~~~~~2~~~~~~~~~0.1~~~~~~~~~~~2~~~~~~~~~~~~~~~~~~~~~~~~~~~~$0.178112$
\\
~~~~1~~~~~~~~~0.1~~~~~~~~~~~~2~~~~~~~~~0.1~~~~~~~~~~~4~~~~~~~~~~~~~~~~~~~~~~~~~~~~$0.641779$
\\ \hline \\
~~~~2~~~~~~~~~~1~~~~~~~~~~~~0.1~~~~~~~~~1~~~~~~~~~~~~1~~~~~~~~~~~~~~~~~~~~~~~~~~~~~$1.25072$
\\
~~~~2~~~~~~~~~~1~~~~~~~~~~~~0.1~~~~~~~~~1~~~~~~~~~~~~2~~~~~~~~~~~~~~~~~~~~~~~~~~~~$0.688701$
\\
~~~~2~~~~~~~~~~1~~~~~~~~~~~~0.1~~~~~~~~~1~~~~~~~~~~~~4~~~~~~~~~~~~~~~~~~~~~~~~~~~~$0.818295$
\\ \hline \\
~~~-3~~~~~~~~~0.1~~~~~~~~~~~0.1~~~~~~~~~3~~~~~~~~~~~~1~~~~~~~~~~~~~~~~~~~~~~~~~~~~~$1.0127$
\\
~~~~3~~~~~~~~~0.1~~~~~~~~~~~0.1~~~~~~~~~3~~~~~~~~~~~~2~~~~~~~~~~~~~~~~~~~~~~~~~~~~$0.610956$
\\
~~~~3~~~~~~~~~0.1~~~~~~~~~~~0.1~~~~~~~~~3~~~~~~~~~~~~4~~~~~~~~~~~~~~~~~~~~~~~~~~~~$0.759971$
\\ \hline \\
~~~-4~~~~~~~~~0.1~~~~~~~~~~~0.1~~~~~~~~~1~~~~~~~~~~~~1~~~~~~~~~~~~~~~~~~~~~~~~~~~~$1.261581$
\\
~~~-4~~~~~~~~~0.1~~~~~~~~~~~0.1~~~~~~~~~1~~~~~~~~~~~~2~~~~~~~~~~~~~~~~~~~~~~~~~~~~$0.688701$
\\
~~~-4~~~~~~~~~0.1~~~~~~~~~~~0.1~~~~~~~~~1~~~~~~~~~~~~4~~~~~~~~~~~~~~~~~~~~~~~~~~~~$0.818295$
\\
\hline\hline\\
\end{tabular}

\end{center}
\begin{center}
Table1c
\begin{tabular}{|l|}
\hline\hline\\ ~~~~~~~~~~~~~~~~~~~~~~~~~~~~~~~~~~~~~~~~~~~~~~~~For $k=-0.5$ (dark energy)\\
\hline\hline
~~~~$\omega$~~~~~~~~~~~$\lambda$~~~~~~~~~~~$\delta$~~~~~~~~~~~$\epsilon$~~~~~~~~~~~$n$
~~~~~~~~~~~~~~~~~~Positive roots ($X_{0}$)
\\
\hline\\
~~~~2~~~~~~~~~~~0.1~~~~~~~~0.1~~~~~~~~~1~~~~~~~~~~~~1~~~~~~~~~~~~~~~~~~~~~~~~~~~~~$0.763465$
\\
~~~~2~~~~~~~~~~~0.1~~~~~~~~0.1~~~~~~~~~1~~~~~~~~~~~~2~~~~~~~~~~~~~~~~~~~~~~~~~~~~~~~~$-$
\\
~~~~2~~~~~~~~~~~0.1~~~~~~~~0.1~~~~~~~~~1~~~~~~~~~~~~4~~~~~~~~~~~~~~~~~~~~~~~~~~~~$0.761803$
\\ \hline \\
~~~~1~~~~~~~~~~~0.1~~~~~~~~~2~~~~~~~~~~1~~~~~~~~~~~~1~~~~~~~~~~~~~~~~~~~~~~~~~~~~~~$0.7107$
\\
~~~~1~~~~~~~~~~~0.1~~~~~~~~~2~~~~~~~~~~1~~~~~~~~~~~~2~~~~~~~~~~~~~~~~~~~~~~~~~~~~~~~~$-$
\\
~~~~1~~~~~~~~~~~0.1~~~~~~~~~2~~~~~~~~~~1~~~~~~~~~~~~4~~~~~~~~~~~~~~~~~~~~~~~~~~~~~$0.510191$
\\ \hline \\
~~~-0.5~~~~~~~~~~3~~~~~~~~~0.1~~~~~~~~0.1~~~~~~~~~~~1~~~~~~~~~~~~~~~~~~~~~~~~~~~~~~$1.70536$
\\
~~~~0.5~~~~~~~~~~3~~~~~~~~~0.1~~~~~~~~0.1~~~~~~~~~~~2~~~~~~~~~~~~~~~~~~~~~~~~~~~~~~~~$-$
\\
~~~~0.5~~~~~~~~~~3~~~~~~~~~0.1~~~~~~~~0.1~~~~~~~~~~~4~~~~~~~~~~~~~~~~~~~~~~~~~~~~~~$0.984339$
\\ \hline \\
~~~-2~~~~~~~~~~~0.1~~~~~~~~0.1~~~~~~~~~3~~~~~~~~~~~~1~~~~~~~~~~~~~~~~~~~~~~~~~~~~~~$0.520046$
\\
~~~-2~~~~~~~~~~~0.1~~~~~~~~0.1~~~~~~~~~3~~~~~~~~~~~~2~~~~~~~~~~~~~~~~~~~~~~~~~~~~~~~~~~~$-$
\\
~~~-2~~~~~~~~~~~0.1~~~~~~~~0.1~~~~~~~~~3~~~~~~~~~~~~4~~~~~~~~~~~~~~~~~~~~~~~~~~~~~~$0.670155$
\\ \hline \\
~~~-3~~~~~~~~~~~0.1~~~~~~~~0.1~~~~~~~~~1~~~~~~~~~~~~1~~~~~~~~~~~~~~~~~~~~~~~~~~~~~~$0.763465$
\\
~~~-3~~~~~~~~~~~0.1~~~~~~~~0.1~~~~~~~~~1~~~~~~~~~~~~2~~~~~~~~~~~~~~~~~~~~~~~~~~~~~~~~~~~$-$
\\
~~~-3~~~~~~~~~~~0.1~~~~~~~~0.1~~~~~~~~~1~~~~~~~~~~~~4~~~~~~~~~~~~~~~~~~~~~~~~~~~~~~$0.761803$
\\ \hline \\
~~~-5~~~~~~~~~~~0.1~~~~~~~~0.1~~~~~~~~~1~~~~~~~~~~~~1~~~~~~~~~~~~~~~~~~~~~~~~~~~~~~$1.04294$
\\
~~~-5~~~~~~~~~~~0.1~~~~~~~~0.1~~~~~~~~~1~~~~~~~~~~~~2~~~~~~~~~~~~~~~~~~~~~~~~~~~~~~~~~~~$-$
\\
~~~-5~~~~~~~~~~~0.1~~~~~~~~0.1~~~~~~~~~1~~~~~~~~~~~~4~~~~~~~~~~~~~~~~~~~~~~~~~~~~~~$0.761803$
\\\hline\hline
 \end{tabular}

\end{center}
\begin{center}
Table1d
\begin{tabular}{|l|}

\hline\hline\\ ~~~~~~~~~~~~~~~~~~~~~~~~~~~~~~~~~~~~~~~~~~~~~~~~For $k=-1$ ($\Lambda$CDM)\\
\hline\hline
~~~~$\omega$~~~~~~~~~~~$\lambda$~~~~~~~~~~~$\delta$~~~~~~~~~~~$\epsilon$~~~~~~~~~~~$n$
~~~~~~~~~~~~~~~Positive roots ($X_{0}$)
\\
\hline\\
~~~~3~~~~~~~~~~~0.1~~~~~~~~0.2~~~~~~~~~~1~~~~~~~~~~1~~~~~~~~~~~~~~~~~~~~~~~~~~~~~~$1.56222$
\\
~~~~3~~~~~~~~~~~0.1~~~~~~~~0.2~~~~~~~~~~1~~~~~~~~~~2~~~~~~~~~~~~~~~~~~~~~~~~~~~~~~~$-$
\\
~~~~3~~~~~~~~~~~0.1~~~~~~~~0.2~~~~~~~~~~1~~~~~~~~~~4~~~~~~~~~~~~~~~~~~~~~~~~~~$0.665843$
\\ \hline \\
~~~~1~~~~~~~~~~~0.1~~~~~~~~0.1~~~~~~~~~~1~~~~~~~~~~1~~~~~~~~~~~~~~~~~~~~~~~~~~~~~$0.522207$
\\
~~~~1~~~~~~~~~~~0.1~~~~~~~~0.1~~~~~~~~~~1~~~~~~~~~~2~~~~~~~~~~~~~~~~~~~~~~~~~~~~~~~$-$
\\
~~~~1~~~~~~~~~~~0.1~~~~~~~~0.1~~~~~~~~~~1~~~~~~~~~~4~~~~~~~~~~~~~~~~~~~~~~~~~~~~$0.686147$
\\ \hline \\
~~~-0.5~~~~~~~~~1~~~~~~~~~~0.1~~~~~~~~~~3~~~~~~~~~~1~~~~~~~~~~~~~~~~~~~~~~~~~~~~$0.981042$
\\
~~~-0.5~~~~~~~~~1~~~~~~~~~~0.1~~~~~~~~~~3~~~~~~~~~~2~~~~~~~~~~~~~~~~~~~~~~~~~~~~$0.367083$
\\
~~~-0.5~~~~~~~~~1~~~~~~~~~~0.1~~~~~~~~~~3~~~~~~~~~~4~~~~~~~~~~~~~~~~~~~~~~~~~~~$0.558615$
\\ \hline \\
~~~-2~~~~~~~~~~0.1~~~~~~~~~0.1~~~~~~~~~~1~~~~~~~~~~1~~~~~~~~~~~~~~~~~~~~~~~~~~~$0.522208$
\\
~~~-2~~~~~~~~~~0.1~~~~~~~~~0.1~~~~~~~~~~1~~~~~~~~~~2~~~~~~~~~~~~~~~~~~~~~~~~~~~$0.582659$
\\
~~~-2~~~~~~~~~~0.1~~~~~~~~~0.1~~~~~~~~~~1~~~~~~~~~~4~~~~~~~~~~~~~~~~~~~~~~~~~~~$0.686145$
\\ \hline \\

~~~-4~~~~~~~~~~0.1~~~~~~~~~0.1~~~~~~~~~~1~~~~~~~~~~1~~~~~~~~~~~~~~~~~~~~~~~~~~~~$0.522216$
\\
~~~-4~~~~~~~~~~0.1~~~~~~~~~0.1~~~~~~~~~~1~~~~~~~~~~2~~~~~~~~~~~~~~~~~~~~~~~~~~~~$0.582659$
\\
~~~-4~~~~~~~~~~0.1~~~~~~~~~0.1~~~~~~~~~~1~~~~~~~~~~4~~~~~~~~~~~~~~~~~~~~~~~~~~~~$0.686147$
\\\hline\hline

\end{tabular}

\end{center}
\begin{center}
Table1e
\begin{tabular}{|l|}
\hline\hline\\ ~~~~~~~~~~~~~~~~~~~~~~~~~~~~~~~~~~~~~~~~~~~~~~~~For $k=-2$ (phantom)\\
\hline\hline
~~~~$\omega$~~~~~~~~~~~$\lambda$~~~~~~~~~~~$\delta$~~~~~~~~~~~$\epsilon$~~~~~~~~~~~$n$
~~~~~~~~~~~~~~~Positive roots ($X_{0}$)
\\
\hline\\
~~~~2~~~~~~~~~~~0.1~~~~~~~~~~0.1~~~~~~~~~1~~~~~~~~~~~1~~~~~~~~~~~~~~~~~~~~~~~~~~$-$
\\
~~~~2~~~~~~~~~~~0.1~~~~~~~~~~0.1~~~~~~~~~1~~~~~~~~~~~2~~~~~~~~~~~~~~~~~~~~~~~~~~$-$
\\
~~~~2~~~~~~~~~~~0.1~~~~~~~~~~0.1~~~~~~~~~1~~~~~~~~~~~4~~~~~~~~~~~~~~~~~~~~~~$1.11744$

\\ \hline \\
~~~~1~~~~~~~~~~~0.1~~~~~~~~~~0.1~~~~~~~~~1~~~~~~~~~~~1~~~~~~~~~~~~~~~~~~~~~~~~~~$-$

\\
~~~~1~~~~~~~~~~~0.1~~~~~~~~~~0.1~~~~~~~~~1~~~~~~~~~~~2~~~~~~~~~~~~~~~~~~~~~~~~~~$-$~

\\
~~~~1~~~~~~~~~~~0.1~~~~~~~~~~0.1~~~~~~~~~1~~~~~~~~~~~4~~~~~~~~~~~~~~~~~~~~~~$1.11744$
\\ \hline \\
~~~-0.5~~~~~~~~~0.1~~~~~~~~~~0.1~~~~~~~~~1~~~~~~~~~~~1~~~~~~~~~~~~~~~~~~~~~~~~~~$-$
\\
~~~-0.5~~~~~~~~~0.1~~~~~~~~~~0.1~~~~~~~~~1~~~~~~~~~~~2~~~~~~~~~~~~~~~~~~~~~~~~~~$-$
\\
~~~-0.5~~~~~~~~~0.1~~~~~~~~~~0.1~~~~~~~~~1~~~~~~~~~~~4~~~~~~~~~~~~~~~~~~~~~~$1.11744$

\\ \hline \\
~~~-2~~~~~~~~~~~0.1~~~~~~~~~~0.1~~~~~~~~~1~~~~~~~~~~~1~~~~~~~~~~~~~~~~~~~~~~~~~~$-$
\\
~~~~2~~~~~~~~~~~0.1~~~~~~~~~~0.1~~~~~~~~~1~~~~~~~~~~~2~~~~~~~~~~~~~~~~~~~~~~~~~~$-$

\\
~~~~2~~~~~~~~~~~0.1~~~~~~~~~~0.1~~~~~~~~~1~~~~~~~~~~~4~~~~~~~~~~~~~~~~~~~~~~$1.11744$
\\ \hline \\
~~~-3~~~~~~~~~~~0.1~~~~~~~~~~~2~~~~~~~~~~1~~~~~~~~~~~1~~~~~~~~~~~~~~~~~~~~~~~~~~$-$

\\
~~~~3~~~~~~~~~~~0.1~~~~~~~~~~~2~~~~~~~~~~1~~~~~~~~~~~2~~~~~~~~~~~~~~~~~~~~~~~~~~$-$
\\
~~~~3~~~~~~~~~~~0.1~~~~~~~~~~~2~~~~~~~~~~1~~~~~~~~~~~4~~~~~~~~~~~~~~~~~~~~~~$1.45782$
\\ \hline \\
~~~-4~~~~~~~~~~~0.1~~~~~~~~~~0.1~~~~~~~~~1~~~~~~~~~~~1~~~~~~~~~~~~~~~~~~~~~~~~~~$-$
\\
~~~~4~~~~~~~~~~~0.1~~~~~~~~~~0.1~~~~~~~~~1~~~~~~~~~~~2~~~~~~~~~~~~~~~~~~~~~~~~~~$-$
\\
~~~~4~~~~~~~~~~~0.1~~~~~~~~~~0.1~~~~~~~~~1~~~~~~~~~~~4~~~~~~~~~~~~~~~~~~~~~~$1.11745$
\\
 \hline\hline
\end{tabular}

\end{center}
~~~~~~~~~~~~~~~{\bf Table 1a-e:} Values of $X_{0}$ for different
values of parameters
$\lambda, \delta, \epsilon, n, \omega$ and $k$.\\
\vspace{.5cm}

\begin{figure}

\includegraphics[height=2in, width=2in]{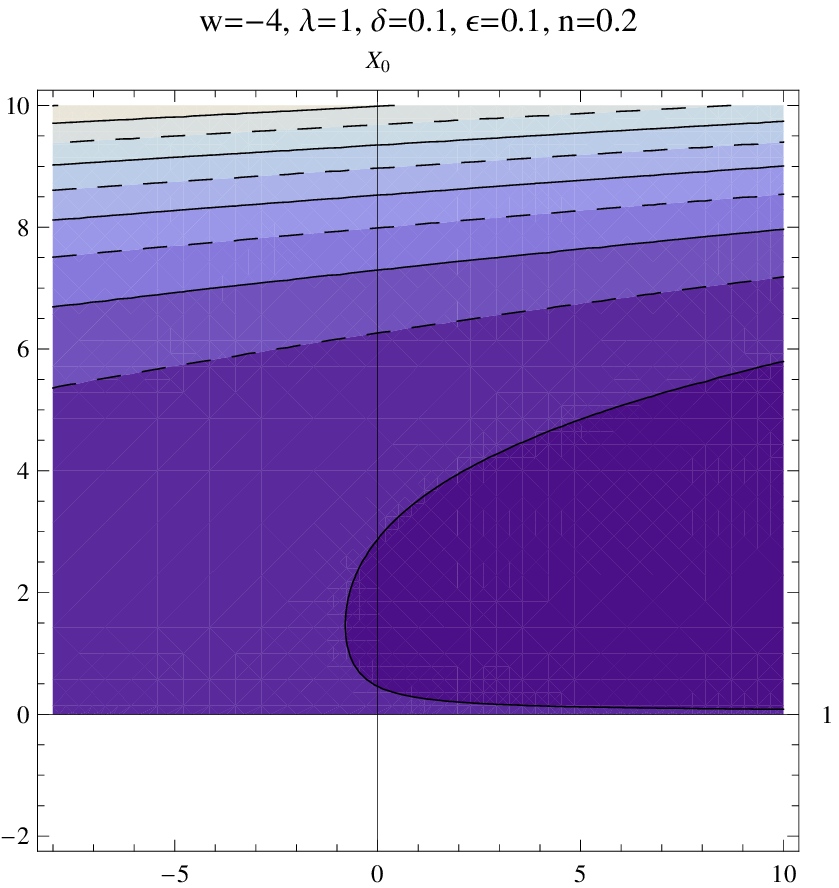}~~~~~~~
\includegraphics[height=2in, width=2in]{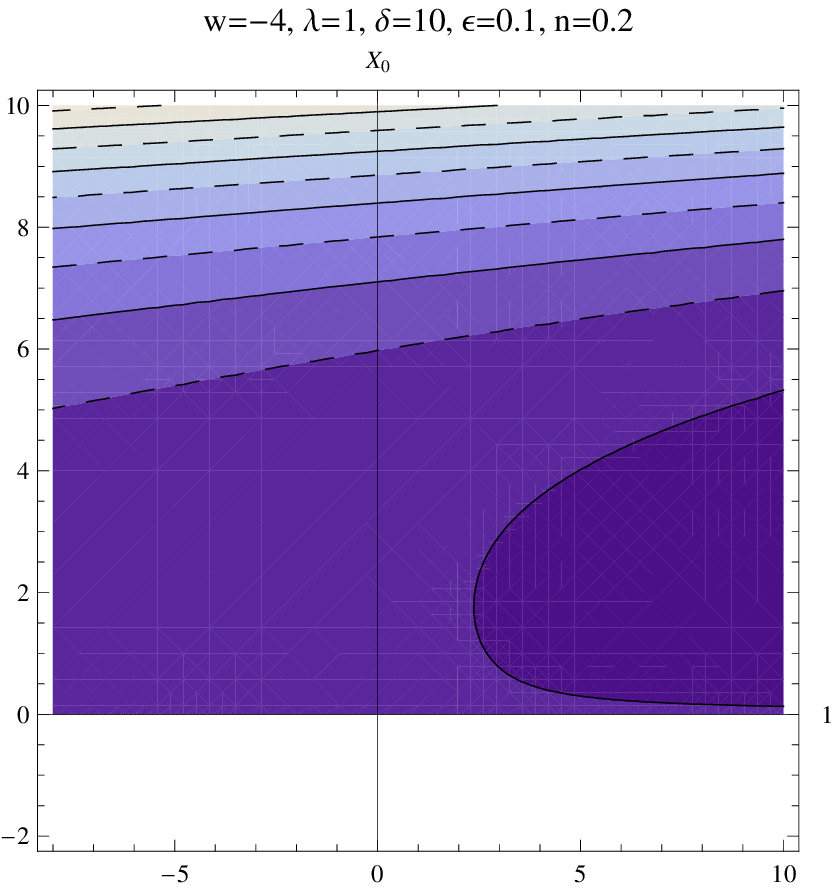}~~~~~~~
\includegraphics[height=2in, width=2in]{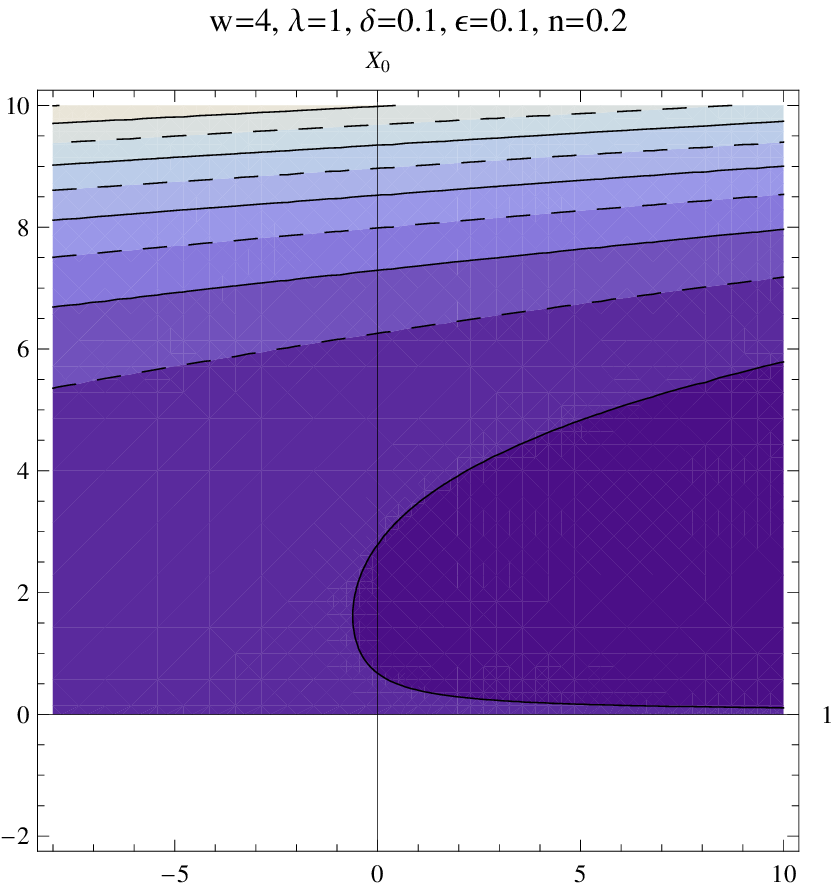}\\
\vspace{1mm}
~~~~~~~~~~~~~~~~~~~Fig.1a~~~~~~~~~~~~~~~~~~~~~~~~~~~~~~~~~~~Fig.1b~~~~~~~~~~~~~~~~~~~~~~~~~~~~~~~~~~~~~~Fig.1c\\
\vspace{3mm}

\includegraphics[height=2in, width=2in]{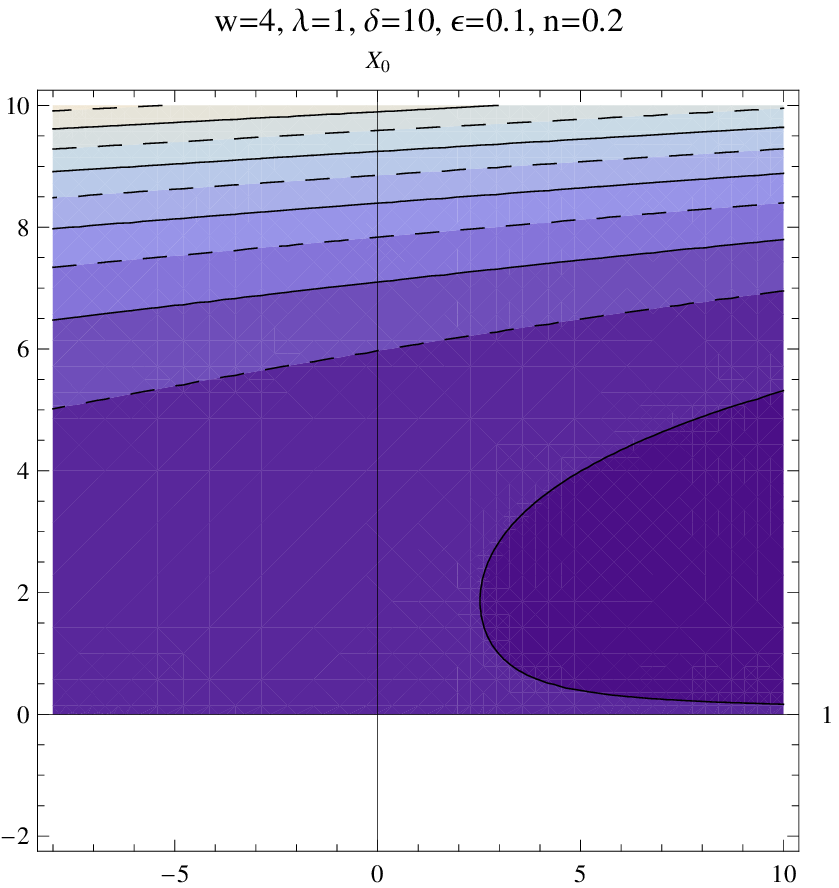}~~~~~~~
\includegraphics[height=2in, width=2in]{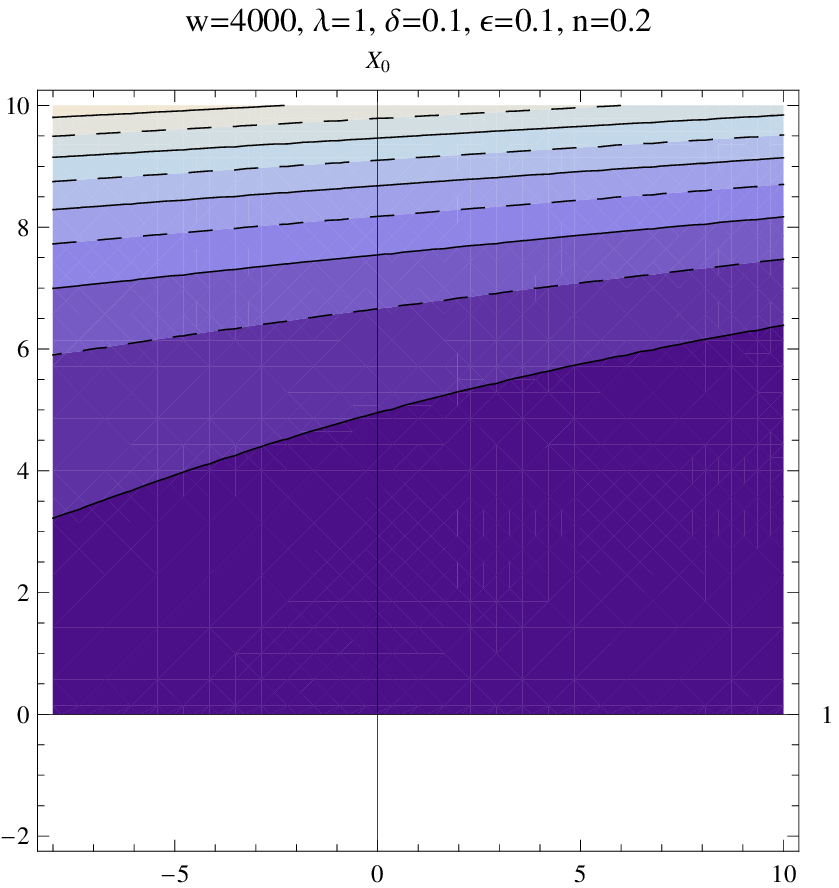}~~~~~~~
\includegraphics[height=2in, width=2in]{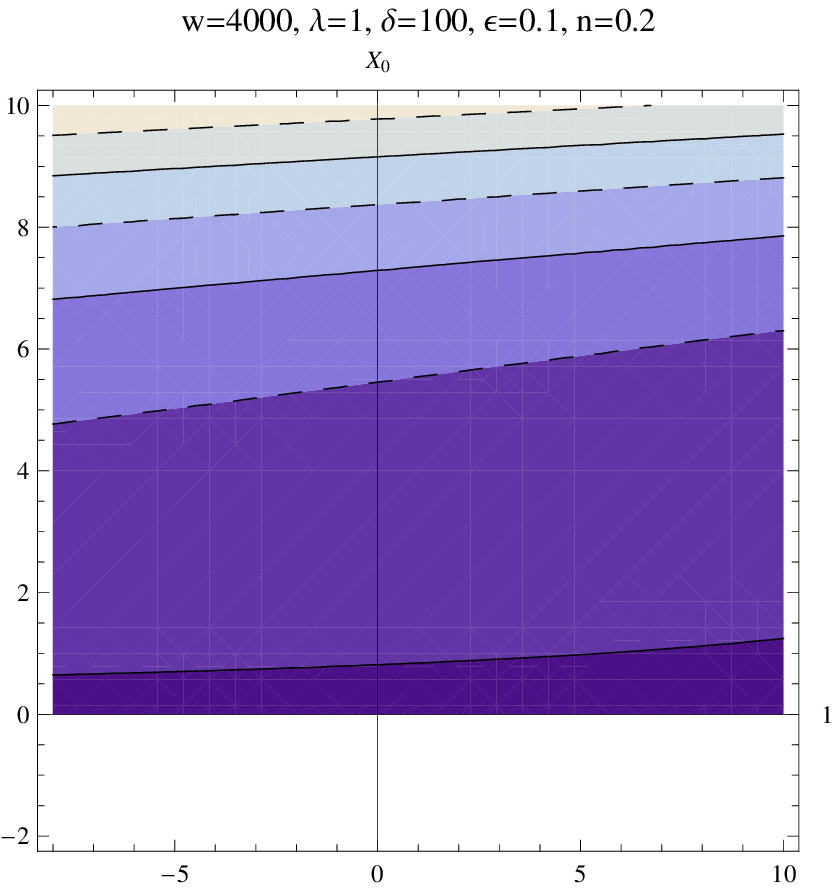}\\
\vspace{1mm}
~~~~~~~~~~~~~~~~~~Fig.1d~~~~~~~~~~~~~~~~~~~~~~~~~~~~~~~~~Fig.1e~~~~~~~~~~~~~~~~~~~~~~~~~~~~~~~~~~~Fig.1f\\
\vspace{3mm}

\includegraphics[height=2in, width=2in]{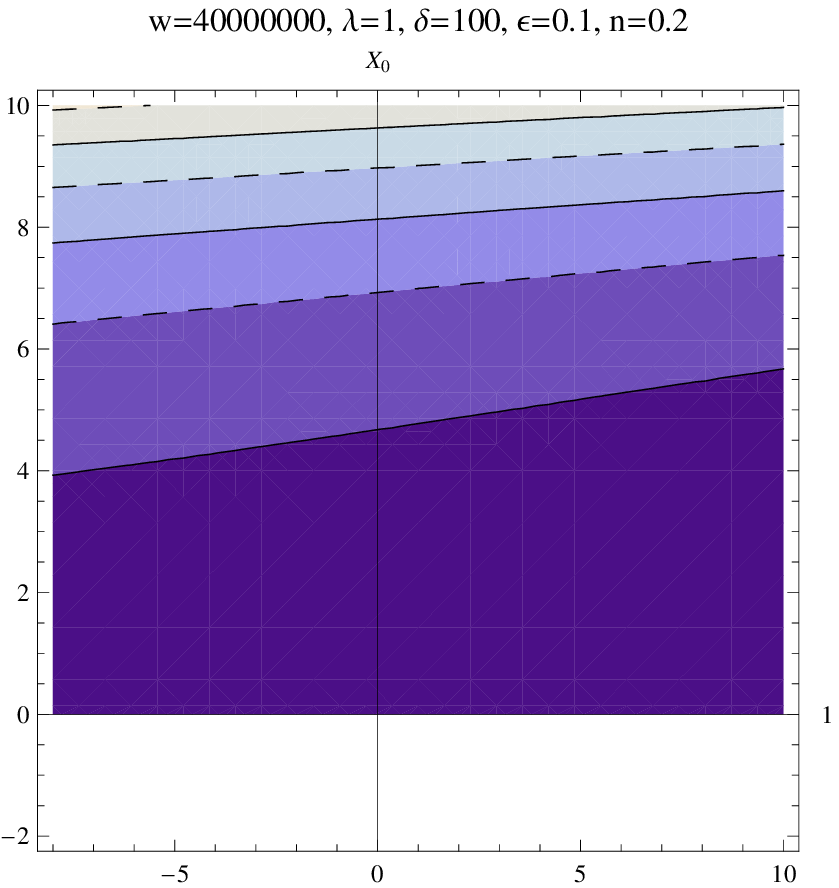}~~~~~~~
\includegraphics[height=2in, width=2in]{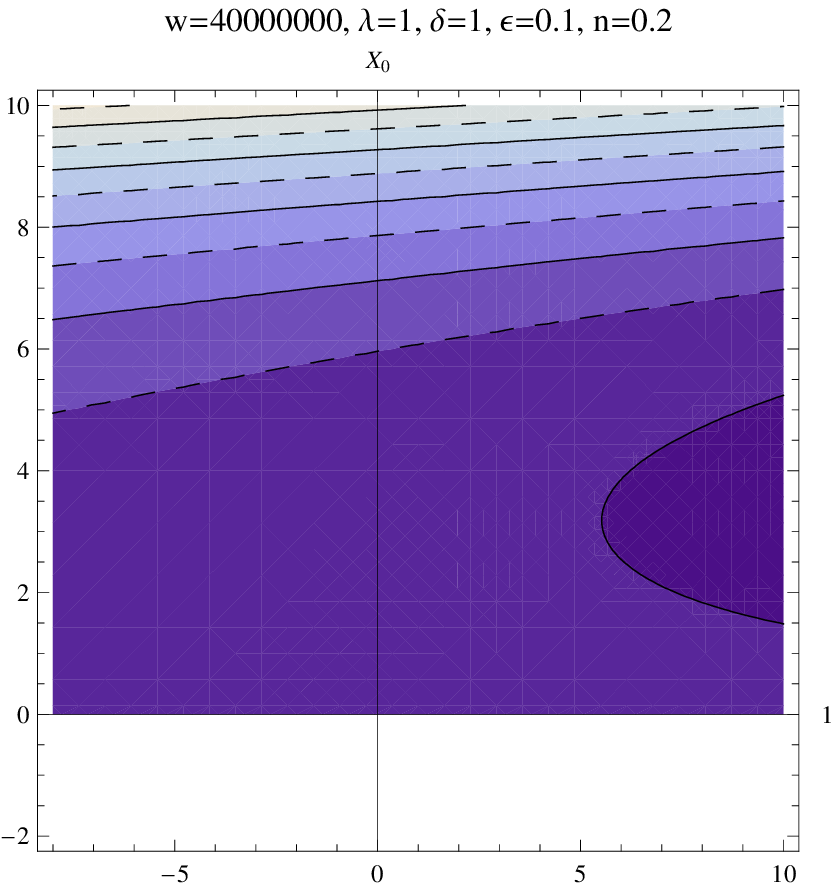}~~~~~~~
\includegraphics[height=2in, width=2in]{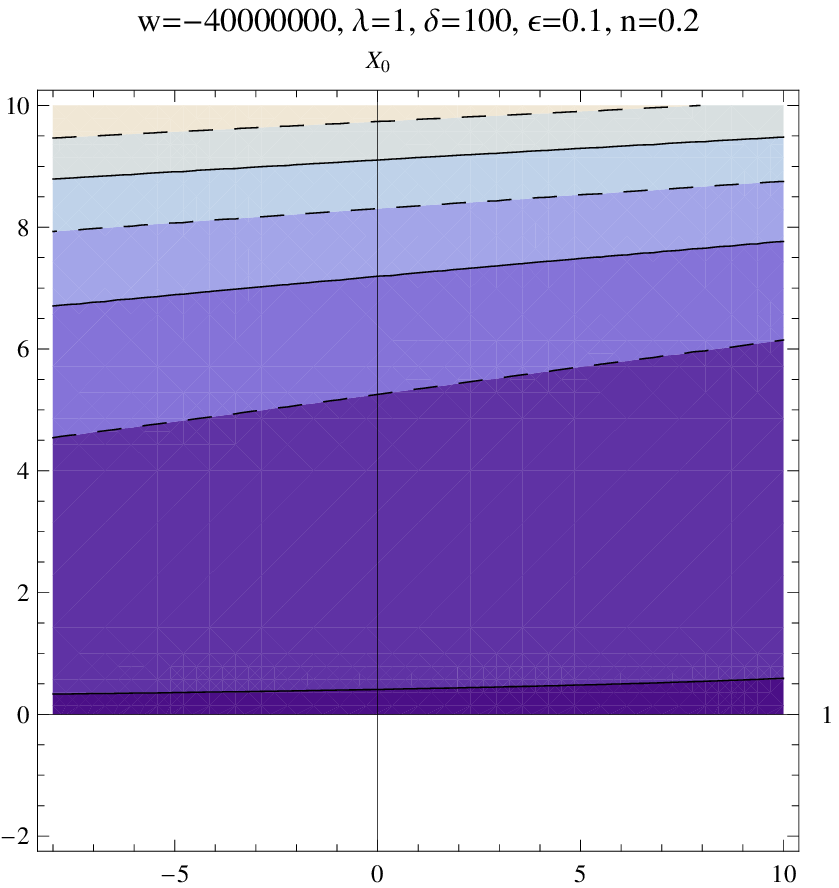}\\
\vspace{1mm}
~~~~~~~~~~~~~~~~~~Fig.1g~~~~~~~~~~~~~~~~~~~~~~~~~~~~~~~~~~~Fig.1h~~~~~~~~~~~~~~~~~~~~~~~~~~~~~~~~~~~Fig.1i\\
\vspace{3mm}

\includegraphics[height=2in, width=2in]{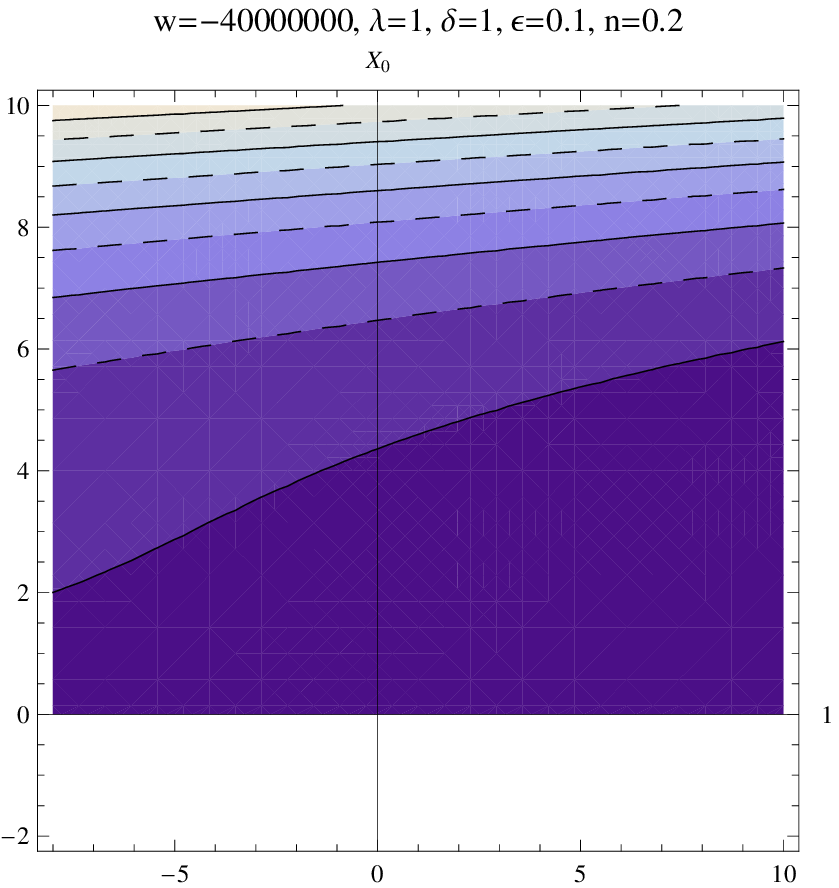}~~~~~~~
\includegraphics[height=2in, width=2in]{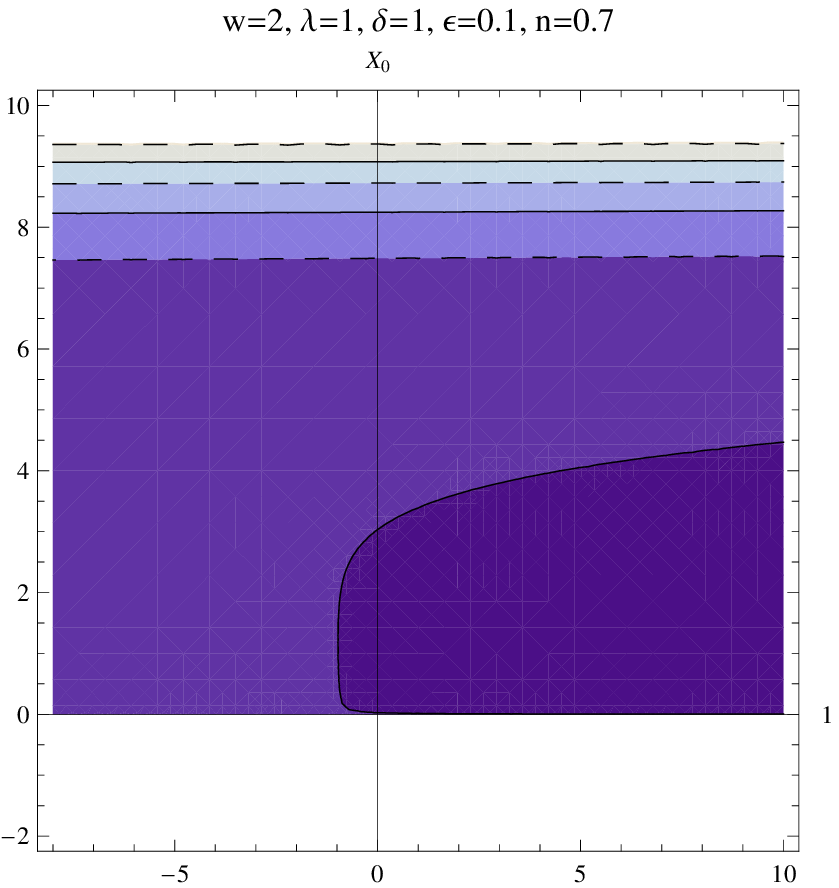}~~~~~~~
\includegraphics[height=2in, width=2in]{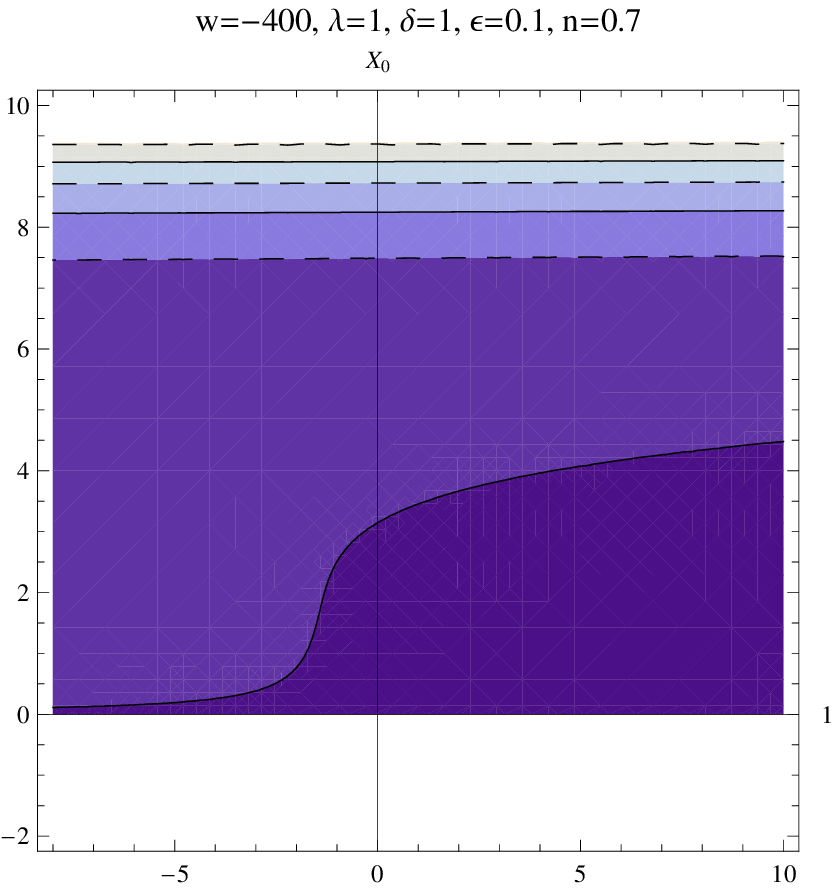}\\
\vspace{1mm}
~~~~~~~~~~~~~~~~~~Fig.1j~~~~~~~~~~~~~~~~~~~~~~~~~~~~~~~~~~~Fig.1k~~~~~~~~~~~~~~~~~~~~~~~~~~~~~~~~~~~Fig.1l\\
\end{figure}
\vspace{3mm}
\begin{figure}
\includegraphics[height=2in, width=2in]{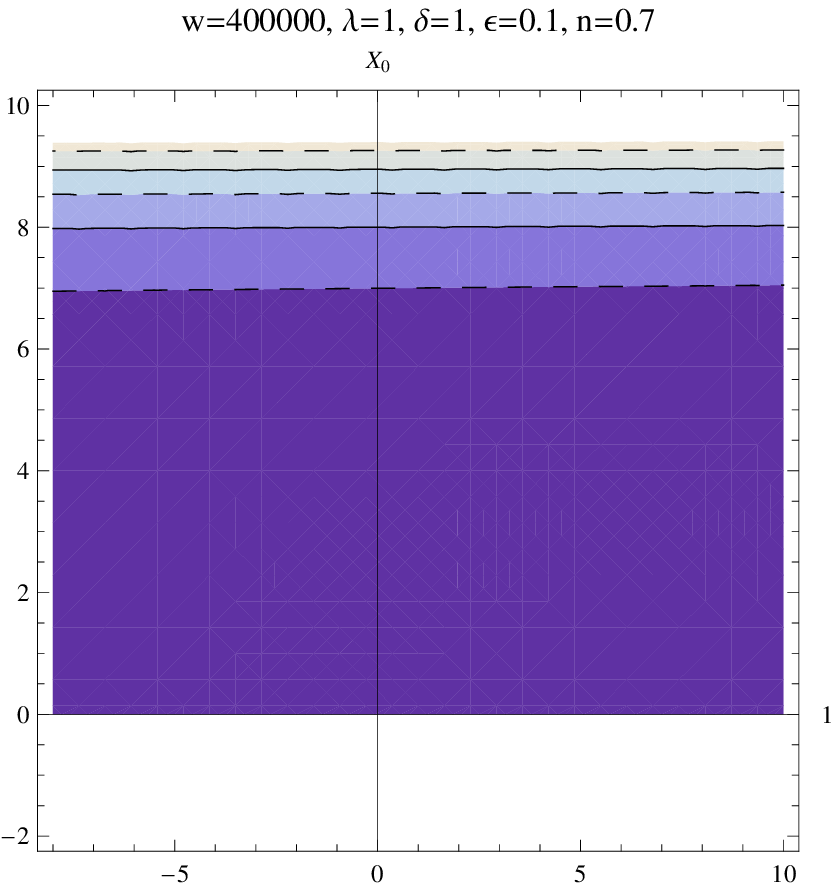}~~~~~~~
\includegraphics[height=2in, width=2in]{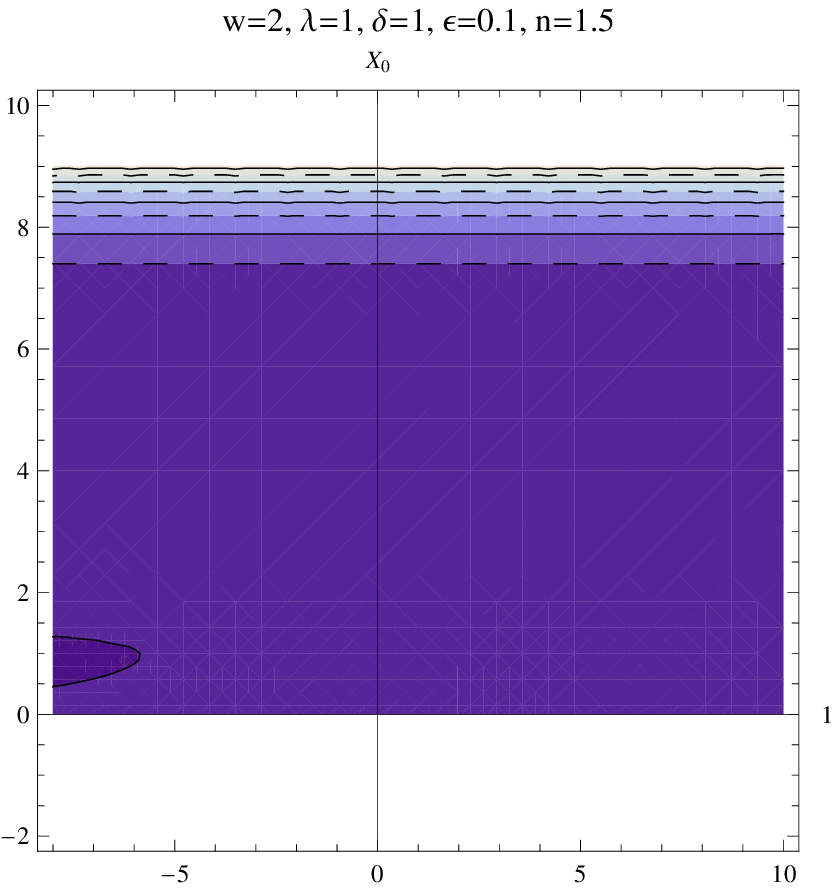}~~~~~~~
\includegraphics[height=2in, width=2in]{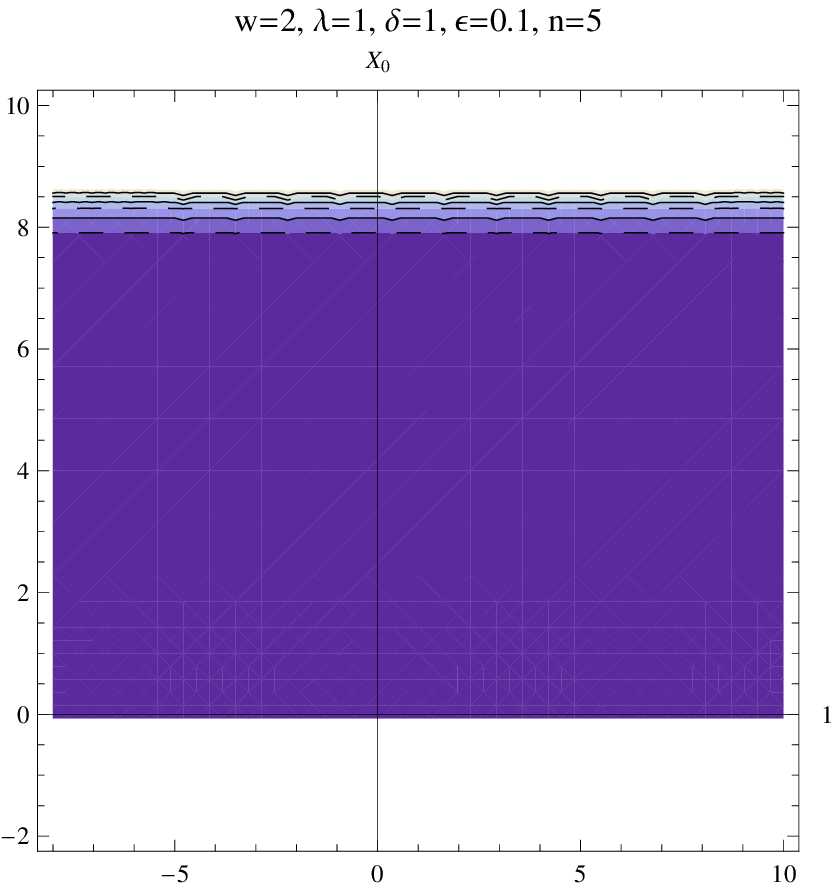}\\
\vspace{1mm}
~~~~~~~~~~~~~~~~~~Fig.1m~~~~~~~~~~~~~~~~~~~~~~~~~~~~~~~~~~~Fig.1n~~~~~~~~~~~~~~~~~~~~~~~~~~~~~~~~~~~Fig.1o\\

\vspace{0.5cm} {\bf Figs. 1a-o} show the $X_0$ contours in the
$k$-$X_0$ plane for different values of $\omega$. \vspace{2cm}
\end{figure}

Looking at the tables, we can see that the result of collapse does
not quite depend on the value of the Galileon parameter $\omega$.
Here, the two important parameters are $n$, and the EoS parameter,
$k$. With a slight variation of all other parameters we have
considered two extremal variations for $k$ and $n$. When the EoS
parameter $k$ is $1$, i.e., stiff perfect fluid, one we get
positive solutions for higher values of $n$, irrespective of the
value of the $\omega$ in the range $-4<\omega<3$. In the limit
$r\rightarrow 0$, i.e. near the singularity, as $n$ increases
$P(r)$ decreases, thus lowering the value of the Galileon scalar
field $\phi$. Apparently, it says lesser the effect of the
Galileon scalar field upon the potential more is the chance to
have a NS. This is contrary to our expectation from our previous
experience of gravitational collapse in Brans-Dicke theory
\cite{Rudra2}. In radiation era ($k=\frac{1}{3}$), naked
singularity is the ultimate fate of the universe irrespective of
the values of the parameters. In accelerating universe ($k=-0.5$)
the trend is almost similar to the radiation era. Except for
$n=2$, the collapse always results in a naked singularity. At
phantom crossing ($k=-1$) we see that the tendency of getting a
naked singularity increases as the value of $\omega$ is decreased.
In phantom era ($k=-2$) contrary to our expectations from our
previous experience \cite{Rudra2} the tendency of occurrence of
naked singularity decreases appreciably. Nevertheless we get naked
singularities for very high values of $n$, i.e. under the
influence of lower values of the Galileon scalar field. As only
two parameters are controlling the end fate, we will plot their
variations in the fig $1a-o$. In figures 1a to 1o we have plotted
the $k-X_0$ contours for increasing values of $\omega$ and $n$.

\section{Graphical Analysis}

\textbf{Contour plots are generally drawn to show the simultaneous
variations of more than one quantity in a 2D plot. In the
algebraic eqn. 28, we can see that there are many parameters. Now
out of these, the main parameters that really control the
collapsing scenario are $n$ and $\omega$ , as determined from the
tables. Since here we are interested in studying the end state of
collapse in different cosmological eras (i.e. for different values
of $k$), we have been inclined to generate $k$ vs $X_{0}$ plot.
The contour lines show the existence of positive solutions of
$X_{0}$ for various values of $k$, for a particular set of values
of the parameters involved. It can be seen that for a particular
set of parametric values, eqn. 28 becomes non-linear and hence
more than one solution exists. In the figures, each contour lines
correspond to one such solution of $X_{0}$. The regions having
identical colours exhibit almost identical properties. It is seen
that, as the value of $n$ is increased, the contour lines get
raised and crowd around higher values of $X_{0}$. So the
dependence on $n$ is quite clear. For relatively higher values of
$n$, there is a greater tendency of getting positive roots of
$X_{0}$, resulting in a naked singularity .}

\section{Conclusion}\label{Discussion}
In this work, first we have assumed the spherically symmetric
space-time model with Vaidya null radiation and perfect fluid in
the background of one of the modified gravity like Galileon
gravity, which is more generalized form of Brans-Dicke gravity. We
have determined the solutions of Einstein equations in Galileon
gravity thus constituting the generalized Vaidya metric. The
solution may be called generalized Vaidya solution in Galileon
gravity theory. Then we investigated the existence of the radial
null geodesic from the final fate of the collapsing object.
Chances of having such geodesic points corresponds to the chances
of having a NS. If not then the construction of a BH is confirmed.
In this paper we see that by proper fine tuning of the parameters
it is always possible to have a uncensored singularity as the fate
of the gravitational collapse in Galileon gravity. We know from
our previous works \cite{Rudra1, Rudra2} that a strong modified
gravity like the Brans-Dicke gravity censors the singularity which
is formed by the gravitational collapse. The role of EoS parameter
$k$ has been analyzed for different eras. When the EoS parameter
$k$ is $1$, i.e., stiff perfect fluid, we get positive solutions
for higher values of $n$, irrespective of the value of the
$\omega$ in the range $-4<\omega<3$, which determines the
possibility of NS. For radiation and dark energy eras, the NS is
the ultimate fate of the gravitational collapse. At phantom
crossing ($k=-1$), we see that the tendency of getting a naked
singularity increases as the value of $\omega$ is decreased. In
phantom era ($k=-2$), the tendency of occurrence of naked
singularity decreases. In figures 1a to 1o we also see that the
natures of $k-X_0$ contours for varying values of two parameters
$\omega$ and $n$. From the plots we get an idea that the
possibility of NS increases for higher values of $n$ and the lower
values of $\omega$. From this we can conclude that the Galileon
gravity, which is an infrared modification of Einstein gravity is
perhaps not as strong a gravity theory as its counterparts.

\begin{contribution}\\\\
{\bf Acknowledgement :}\\

Authors thank the anonymous referee for his/her invaluable
comments that helped to identify the flaws and to improve the
quality of the manuscript.

\end{contribution}

\end{document}